\documentclass[11pt,psamsfonts]{amsart}
\usepackage{amsmath}
\usepackage{amsthm}
\usepackage{amssymb}
\usepackage{amscd}
\usepackage{amsfonts}
\usepackage{amsbsy}
\usepackage{booktabs}
\usepackage{calc}
\usepackage{caption}
\usepackage{color}
\usepackage{graphicx}
\usepackage{lineno}
\usepackage{lscape}
\usepackage[misc]{ifsym}
\usepackage{multirow}
\usepackage{setspace}
\usepackage{subfigure}

\usepackage{geometry}
\begin{document}
\title[]
{Revisiting the distributions of Jupiter's irregular moons: II. orbital characteristics}
 \author{Fabao~Gao$^{1,2}$, Xia Liu$^{1}$}
 \address{${}^1$School of Mathematical Science, Yangzhou University, Yangzhou 225002, China}
  \address{${}^2$ Departament de Matem$\grave{\text{a}}$tiques, Universitat Aut$\grave{\text{o}}$noma de Barcelona, Bellaterra 08193, Barcelona, Catalonia, Spain}
  \address{\textnormal{E-mails: gaofabao@sina.com ( \Letter\ Fabao Gao, ORCID 0000-0003-2933-1017), liuxiayzu@sina.com (Xia Liu)}}


\begin{abstract}
This paper statistically describes the orbital distribution laws of Jupiter's irregular moons, most of which are members of the Ananke, Carme and Pasiphae groups. By comparing 19 known continuous distributions, it is verified that suitable distribution functions exist to describe the orbital distributions of these natural satellites. For each distribution type, interval estimation is used to estimate the corresponding parameter values. At a given significance level, a one-sample Kolmogorov-Smirnov non-parametric test is applied to verify the specified distribution, and we often select the one with the largest $p$-value. The results show that the semi-major axis, mean inclination and orbital period of the moons in the Ananke group and Carme group obey Stable distributions. In addition, according to Kepler's third law of planetary motion and by comparing the theoretically calculated best-fitting cumulative distribution function (CDF) with the observed CDF, we demonstrate that the theoretical distribution is in good agreement with the empirical distribution. Therefore, these characteristics of Jupiter's irregular moons are indeed very likely to follow some specific distribution laws, and it will be possible to use these laws to help study certain features of poorly investigated moons or even predict undiscovered ones.
\end{abstract}

\maketitle

\section{Introduction}
The giant Jupiter system is often referred to as a miniature solar system \cite{Morbidelli}. Jupiter's gravity is strong enough to keep objects in orbit over 0.2 AU away, which means that there is a particularly large space around Jupiter for researchers to examine, possibly hiding natural satellites that have not yet been discovered. Since the composition of Jupiter is similar to that of the Sun, the exploration of Jupiter can help to gain a deeper understanding of the solar system. The moons of Jupiter are divided into regular and irregular moons. Irregular moons are characterized by a high eccentricity and inclination, which are distinct from the near-circular, uninclined orbits of regular moons. These distant retrograde moons can be grouped into at least three main orbital groupings and are considered the remnants of three once-larger parent bodies that were broken apart during collisions with asteroids, comets or other natural satellites \cite{Nesvorny}. These three groups are Ananke, Carme and Pasiphae. The specific classifications of Jupiter's irregular moons can be found in Table 7 in Appendix A, where `current' corresponds to the current discovery that Jupiter has 79 moons and its latest classification and `previous' refers to Jupiter's 69 moons and their previous classifications prior to July 2018.  

Carruba et al.\,\cite{Carruba} integrated orbits of a variety of hypothetical Jovian moons on a long timescale and found that the Lidov-Kozai effect due to solar perturbations plays the most prominent role in secular orbital evolution. Recently, Aschwanden\,\cite{Aschwanden} interpreted the observed quasi-regular geometric patterns of planet or moon distances in terms of a self-organizing system. Researchers must be curious as to whether there are specific rules for the irregular moons of Jupiter. When the total number of Jupiter's moons was still 69, based on the classification of Sheppard and Jewitt\,\cite{Sheppard-Jewitt}, Gao et al.\,\cite{Gao} investigated the distributions of orbital and physical characteristics of Jupiter's moons by using a one-sample Kolmogorov-Smirnov (K-S) non-parametric test (please see \cite{Hogg} for details). Several features of Jupiter's moons have been found to obey logistic distribution and $t$ location-scale distribution. In addition, they verified that the distribution results were helpful in predicting some characteristics of the moons that have not been well studied. Moreover, they also believed that if future observations will allow an increase in the number of Jupiter's moons, the distribution laws may be slightly different, but it will not change significantly over a long period of time.

In recent years, the number of known irregular moons has greatly increased with the powerful observations\,\cite{Jewitt}. It is worth mentioning that the Carnegie Institution for Science announced the discovery of 12 new moons of Jupiter in July 2018\,\cite{Sheppard-Williams}. In addition to the temporary designation of 2 moons in June 2017, 10 of these newly discovered moons are part of the outer group moons that orbit the Jupiter in retrograde orbits. This exciting and important discovery has increased the total number of Jovial satellites to 79\,\cite{Sheppard-Williams, Witze}. Considering that the newly discovered Jupiter's moons are small and a few moons have been recategorized (please see Table 7 in Appendix A for details), do their orbital characteristics still follow certain potential laws? 

In this paper, according to the updated data of the moons of Jupiter, we continue to study the distribution laws of their orbital characteristics, including those of the semi-major axis, inclination, eccentricity, argument of periapsis, longitude of the ascending node and period. Except for the period, the remaining five orbital elements are commonly used to specify an orbit. The semi-major axis and eccentricity determine the shape and size of the orbit. The longitude of ascending node and inclination define the orientation of the orbital plane in which the ellipse is embedded, and the argument of periapsis is the angle from the ascending node to the periapsis, measured in the direction of motion.

By using the one-sample K-S test method in statistics, we verify the dozens of commonly used distributions one by one and calculate the $p$-values corresponding to these distributions. For the same orbital feature, the distribution corresponding to the largest $p$-values is theoretically the one we are looking for, and the closer the $p$-value is to 1, the more likely that the distribution is correct. To describe these distributions analytically, we also calculate the values of the parameters corresponding to these distributions and the confidence intervals corresponding to these parameters through statistical inference. Moreover, the relationship between some orbital characteristics can be further analysed to verify whether the theoretical results of the statistical the prediction are valid. For example, the nonlinear relationship between the semi-major axis ($a$) and the orbital period ($T$) can be expressed in accordance with Kepler's third law of planetary motion. We can infer from the observational data that $a$ obeys distribution $d1$, and $T$ obeys distribution $d2$; however, the distribution of $a$ can also be calculated analytically based on Kepler's third law of planetary motion and distribution $d2$. The distribution of $a$ obtained by analytical calculation is recorded here as $d3$. Therefore, for the orbital characteristics of $a$, the rationality of data inference can be verified by comparing distribution $d1$, which is inferred by the K-S test method, and distribution $d3$, which is obtained by analytical calculation.

\section{Distribution inference based on different moons' groups}

\subsection{Ananke group}
The Ananke group had only 11 moons as of a few months ago but currently consists of 19 moons, including the three most recently discovered moons S/2017 J7, S/2017 J3 and S/2017 J9, and five moons formerly in the Pasiphae group: Euporie, Orthosie, Helike, S2003 J18 and S/2016 J1 (see Appendix A the recategorizations of Jupiter's moons regrouping).

Based on the one-sample K-S test and the orbital characteristics of irregular moons (see Appendix B for more details), we obtain the best-fitting distribution of these orbital features, which are shown in Table 1 (More details can be found in Appendix C). In the fourth column of the table, the confidence interval shows that the true value of these parameters falls close to the measurement with a certain probability. Although the distribution parameter values can be calculated from the observed data, whether the null hypothesis be rejected is greatly influenced by the significance level (usually set to 0.05 or 0.01), so we study the distributions of the orbital features by using $p$-values compared with pre-determined significance levels. If the $p$-value is greater than 0.05, we fail to reject the null hypothesis; otherwise it is rejected. However, if the $p$-values of several distributions corresponding to the same orbital characteristics are greater than 0.05, we try to choose the distribution with the largest $p$-value, which is slightly one-sided but is reasonable, because one event often occurs with a greater probability than the others.

It can be seen from Table 1 that the semi-major axis, the mean inclination and the period all obey Stable distributions (See \cite{Nolan, Kateregga} for more details), and the $p$-values are all greater than 0.9, which is much larger than the commonly used thresholds of 0.05 and 0.01. Since most of the probability density functions (PDFs) of Stable distribution have no closed-form expressions except for a few special cases, they are conveniently represented by a characteristic function (CF). If a random variable has a PDF, then the CF is a Fourier transform of the PDF. Therefore, the CF provides the basis for an alternative path to analyse the results compared to directly using the PDF. The 
relationship between the CF and PDF can be expressed by the following formula:
\begin{equation}
f_X(x)=\frac{1}{2\pi}\int_{-\infty }^{\infty }\phi_X(t)e^{-itx}\mbox{d}t,
\end{equation}
where $f_{X}$ and $\phi _{X }$ are the PDF and CF, respectively, of a random variable $X$. 

A random variable $X$ is called Stable (\cite{Nolan}-\cite{Johannes}) if its CF can be written as
\begin{equation}
\phi_X(t;\alpha,\beta,c,\mu)=\exp\left ( it\mu- |ct|^{\alpha}  (1-i\beta\mbox{sgn}(t)\Phi )\right ),\ \ \ t\in R,
\end{equation}
where $\alpha\in(0,2]$ is the characteristic exponent responsible for the shape of the distribution, $\beta\in[-1,1]$ is the skewness of the distribution and is used to measure asymmetry ($\beta$=0 means symmetrical), $c\in(0,+\infty)$ is the scale parameter, which narrows or extends the distribution around, $\mu\in R$ is the location parameter that shifts the distribution to the left or the right, $\mbox{sgn}(t)$ is the usual sign function and
\begin{equation*}  
\Phi=\left\{  
             \begin{array}{lr}  
             \tan\left(\frac{\pi\alpha}{2}\right), & \alpha\neq1,\\  
             -\frac{2}{\pi}\log|t|, & \alpha=1. 
             \end{array}  
\right.  
\end{equation*}

If $\alpha=0.5$ and $\beta=1$, then $X$ follows a Levy distribution. If $\alpha=1$ or $\alpha=2$, it follows a Cauchy or Gaussian distribution, respectively. 

\begin{table*}[!htb]
\newcommand{\tabincell}[2]{\begin{tabular}{@{}#1@{}}#2\end{tabular}}
\centering
\caption{\label{opt}Inference of the distribution of each orbital characteristics in the Ananke group}
\footnotesize
\rm
\centering
\begin{tabular}{@{}*{7}{l}}
 \toprule
\textbf{Characteristic}&\textbf{Distribution}&\textbf{Parameters}&\textbf{Confidence Intervals}&\textbf{$p$-value}\\
 \toprule

Semi-major  axis(km)&Stable&\tabincell{l}{$\alpha=1.32899$,\\$\beta =-1$,\\$c=211134$,\\$\mu =2.10713\ast 10^{7}$}&\tabincell{l}{$\alpha \epsilon[0,2]$,\\$\beta\epsilon [-1,1]$,\\$c\epsilon[0,Inf]$,\\$\mu\epsilon[-Inf,Inf]$}&\tabincell{l}{0.9118}\\
\hline

Mean  inclination(deg)&Stable&\tabincell{l}{$\alpha=1.42053 $,\\$\beta=-0.205584 $,\\$c=1.62162$,\\$\mu=148.744$}&\tabincell{l}{$\alpha\epsilon[0.746773,2]$,\\$\beta\epsilon[-1,0.999048]$,\\$c\epsilon [0.8027,2.44053]$,\\$\mu\epsilon[147.487, 150.002]$}&\tabincell{l}{0.9109}\\
\hline

Mean  eccentricity&Extreme Value&\tabincell{l}{$\mu=0.235049 $,\\$\sigma=0.0444536 $}&\tabincell{l}{$\mu\epsilon[0.213908,0.256189]$,\\$\sigma\epsilon[0.0318622, 0.0620209]$}&\tabincell{l}{0.4945}\\
\hline

Argument  of periapsis&Loglogistic&\tabincell{l}{$\alpha=5.10234 $,\\$\beta=0.374292  $}&\tabincell{l}{$\alpha\epsilon[4.79185,5.41282]$,\\$\beta\epsilon[0.258487, 0.541979]$}&\tabincell{l}{0.6543}\\
\hline

\tabincell{l}{Longitude of the \\ascending node}&\tabincell{l}{Generalized \\Extreme Value}&\tabincell{l}{$k=-0.784386$,\\$\sigma=127.426$,\\$\mu=188.251$}&\tabincell{l}{$k\epsilon[-1.1982,-0.370575]$,\\$\sigma\epsilon[79.4528,204.364]$,\\$\mu\epsilon[124.222,252.28]$}&\tabincell{l}{0.7088}\\
\hline

Period(days)&Stable&\tabincell{l}{$\alpha=1.20629 $,\\$\beta=-1$,\\$c=8.28652 $,\\$\mu=624.981$}&\tabincell{l}{$\alpha\epsilon[0,2]$,\\$\beta\epsilon[-1,1]$,\\$c\epsilon[0,Inf]$,\\$\mu\epsilon[-Inf,Inf]$}&\tabincell{l}{0.9878}\\
 \toprule

\end{tabular}
\end{table*}

\begin{table*}[!htb]
\newcommand{\tabincell}[2]{\begin{tabular}{@{}#1@{}}#2\end{tabular}}
\centering
\caption{\label{opt1}Inference of the distribution of each orbital characteristics in the Carme group}
\footnotesize
\rm
\begin{tabular}{@{}*{7}{l}}
 \toprule
\textbf{Characteristic}&\textbf{Distribution}&\textbf{Parameters}&\textbf{Confidence Intervals}&\textbf{$p$-value}\\
 \toprule

Semi-major  axis(km)&Stable&\tabincell{l}{$\alpha = 0.987755$,\\$\beta =0.0533906$,\\$c=112441$,\\$\mu =2.32477\ast 10^{7}$}&\tabincell{l}{$\alpha \epsilon [0,2]$,\\$\beta \epsilon[-1,1]$,\\$c\epsilon[0,Inf]$,\\$\mu \epsilon[-Inf,Inf]$}&\tabincell{l}{0.9534}\\
\hline

Mean  inclination(deg)&Stable&\tabincell{l}{$\alpha = 0.835022 $,\\$\beta =-0.329864$,\\$c=0.242389$,\\$\mu = 165.084$}&\tabincell{l}{$\alpha \epsilon [0.424668,1.24538]$,\\$\beta \epsilon[-0.948243,0.288516]$,\\$c\epsilon[0.147881,0.336898]$,\\$\mu \epsilon[164.949, 165.219]$}&\tabincell{l}{0.9749}\\
\hline

Mean  eccentricity&Stable&\tabincell{l}{$\alpha = 1.27463  $,\\$\beta =0.000476987$,\\$c=0.0119451$,\\$\mu = 0.256685 $}&\tabincell{l}{$\alpha \epsilon [0.66043,1.88884]$,\\$\beta \epsilon[-1,1]$,\\$c\epsilon[0.00554199,0.0183482]$,\\$\mu \epsilon[0.248013, 0.265357]$}&\tabincell{l}{0.9996}\\
\hline

Argument  of periapsis&\tabincell{l}{Generalized \\Extreme Value}&\tabincell{l}{$k=-0.459188$,\\$\sigma=119.284$,\\$\mu=149.7$}&\tabincell{l}{$k\epsilon[-0.987774, 0.0693978]$,\\$\sigma \epsilon[76.1928,186.747]$,\\$\mu\epsilon[86.0337, 213.366]$}&\tabincell{l}{0.8058}\\
\hline

\tabincell{l}{Longitude of the \\ascending node}&Loglogistic&\tabincell{l}{$\alpha = 5.00721 $,\\$\beta =0.477994 $}&\tabincell{l}{$\alpha \epsilon [4.62636,5.38805]$,\\$\beta \epsilon[0.330581, 0.6911419]$}&\tabincell{l}{0.7756}\\
\hline

Period(days)&Stable&\tabincell{l}{$\alpha = 0.940971 $,\\$\beta =0.163918$,\\$c=6.45023  $,\\$\mu =724.273$}&\tabincell{l}{$\alpha \epsilon [0,2]$,\\$\beta \epsilon[-1,1]$,\\$c\epsilon[0,Inf]$,\\$\mu\epsilon[-Inf,Inf]$}&\tabincell{l}{0.9936}\\
 \toprule

\end{tabular}
\end{table*}

\begin{table}[!htb]
\newcommand{\tabincell}[2]{\begin{tabular}{@{}#1@{}}#2\end{tabular}}
\centering
\caption{\label{opt2}Inference of the distribution of each orbital characteristics in the Pasiphae group}
\footnotesize
\rm
\begin{tabular}{@{}*{7}{l}}
 \toprule
\textbf{Characteristic}&\textbf{Distribution}&\textbf{Parameters}&\textbf{Confidence Intervals}&\textbf{$p$-value}\\
 \toprule

Semi-major  axis(km)&Extreme Value&\tabincell{l}{$\mu =2.38737\ast 10^{7} $,\\$\sigma =493901 $}&\tabincell{l}{$\mu \epsilon[2.36096e\ast 10^{7},2.41378e\ast 10^{7}]$,\\$\sigma \epsilon[334031, 730286]$}&\tabincell{l}{0.9956}\\
\hline

Mean  inclination(deg)&Normal&\tabincell{l}{$\mu =151.213  $,\\$\sigma =4.33208 $}&\tabincell{l}{$\mu \epsilon [148.814,153.612]$,\\$\sigma \epsilon[3.17163, 6.83213]$}&\tabincell{l}{0.9989}\\
\hline

Mean  eccentricity&Birnbaum-Saunders&\tabincell{l}{$\beta =0.33437 $,\\$\gamma  =0.24734 $}&\tabincell{l}{$\beta \epsilon[0.292838,0.375902]$,\\$\gamma \epsilon[0.158832, 0.335847]$}&\tabincell{l}{0.7280}\\
\hline

Argument  of periapsis&\tabincell{l}{Generalized \\Extreme Value}&\tabincell{l}{$k=0.829357$,\\$\sigma=50.6921$,\\$\mu=87.4714$}&\tabincell{l}{$k\epsilon[-0.171729,1.83044]$,\\$\sigma \epsilon[23.1042,111.222]$,\\$\mu\epsilon[51.4714, 123.471]$}&\tabincell{l}{0.5742}\\
\hline

\tabincell{l}{Longitude of the \\ascending node}&Stable&\tabincell{l}{$\alpha = 0.759928 $,\\$\beta =-1$,\\$c=18.0216$,\\$\mu =309.064 $}&\tabincell{l}{$\alpha \epsilon [0,2]$,\\$\beta \epsilon[-1,1]$,\\$c\epsilon[0,Inf]$,\\$\mu \epsilon[-Inf,Inf]$}&\tabincell{l}{0.9799}\\
\hline

Period(days)&\tabincell{l}{Generalized \\Extreme Value}&\tabincell{l}{$k=-0.460585$,\\$\sigma=26.9631$,\\$\mu=729.292$}&\tabincell{l}{$k\epsilon[-0.825924,-0.095246]$,\\$\sigma \epsilon[17.6755,41.1308]$,\\$\mu\epsilon[ [714.26, 744.325]]$}&\tabincell{l}{0.9997}\\
 \toprule
\end{tabular}
\end{table}

\begin{table}[!htb]
\newcommand{\tabincell}[2]{\begin{tabular}{@{}#1@{}}#2\end{tabular}}
\centering
\caption{\label{opt3}Distribution inference summary}
\footnotesize
\rm
\centering
\begin{tabular}{@{}*{7}{l}}
 \toprule
\textbf{Characteristic}&\textbf{Ananke group}&\textbf{Carme group}&\textbf{Pasiphae group}\\
 \toprule
 Semi-major axis(km)&Stable&Stable&Extreme Value\\
Mean inclination(deg)&Stable&Stable&Normal\\
Mean eccentricity&Extreme Value&Stable&Birnbaum-Saunders\\
Argument of periapsis&Loglogistic&Generalized Extreme Value&Generalized Extreme Value\\
Longitude of the ascending node&Generalized Extreme Value&Loglogistic&Stable\\
Period(days)&Stable&Stable&Generalized Extreme Value\\
 \toprule
\end{tabular}
\end{table}

For the sake of simplicity, the distribution types of other orbital features are briefly introduced (see Tables 1-3 for details). The mean eccentricity obeys an Extreme Value distribution with location parameter $\mu$ and scale parameter $\sigma$. As the value of $\sigma$ increases, the density function curve disperses gradually. The mean and variance of the Extreme Value distribution are $\mu +\nu \sigma $ and $\pi ^{2}\sigma ^{2}/6$, respectively, and here, $\nu$ is the Euler constant. The argument of periapsis obeys a Loglogistic distribution, where $\alpha$ is scale parameter and it is also the median of the distribution. The parameter $\beta>0$ is the shape parameter. The distribution is unimodal when $\beta >1$, and its dispersion decreases as $\beta $ increases. The longitude of the ascending node obeys a Generalized Extreme Value distribution with shape parameter $k$, scale parameter $\sigma$, and location parameter $\mu$.

\subsection{Carme group}
There are 20 moons in the Carme group, of which S/2017 J2, S/2017 J5, and S/2017 J8 are the latest discoveries. S/2003 J19 and S/2011 J1 were not part of the Carme group previously but are now in the Carme group. The semi-major axis, the mean inclination, the eccentricity and the period are subject to Stable distributions. The argument of periapsis obeys a Generalized Extreme Value distribution. The longitude of the ascending node follows a Loglogistic distribution (see Table 2 for details). 

\subsection{Pasiphae group}
To date, 15 moons compose the Pasiphae group, including S/2017 J6, which was also newly discovered. The orbital characteristics in this group obey a distribution significantly different from the other two groups in Tables 1 and 2. Specifically, the semi-major axis follows an Extreme Value distribution, while the corresponding best-fitting distribution in the Ananke group and the Carme group is a Stable distribution. The mean inclination follows a Normal distribution. The mean eccentricity obeys a Birnbaum-Saunders distribution. Both the argument of periapsis and period follow Generalized Extreme Value distributions. The longitude of the ascending node follows a Stable distribution.

\section{Comparison of the Previous and Current Orbital-Property Distributions}

As seen from Table 5, the mean inclination obeys a Stable distribution with a $p$-value of 0.9749. However, the previous best-fitting distribution is a $t$ location-scale distribution with a $p$-value of only 0.6662. Similarly, the optimal distribution of the mean eccentricity is a Stable distribution with a $p$-value of 0.9996. All the optimal distributions of these three orbital elements are Stable distributions, which may indicate that the moons in the Carme group are likely to have the same origin; that is, they may have been created from the split of the same parent asteroid.

\begin{table}[]
 \renewcommand{\arraystretch}{1}
\newcommand{\tabincell}[2]{\begin{tabular}{@{}#1@{}}#2\end{tabular}}
\centering
\caption{\label{opt4}Distributions of current and previous orbital characteristics in the Carme group}
\footnotesize
\scalebox{0.8}{
\begin{tabular}{@{}*{7}{l}}
 \toprule
&&\textbf{Current distribution}&&&\textbf{Previous distribution}\\
 \toprule
\textbf{Characterastic}&\textbf{Distribution}&\textbf{Parameters}&\textbf{$p$-value}&\textbf{Distribution}&\textbf{Parameters}&\textbf{$p$-value}\\
 \toprule

Semi-major  axis(km)&Stable&\tabincell{l}{$\alpha = 0.987755$,\\$\beta =0.0533906$,\\$c=112441$,\\$\mu =2.32477\ast 10^{7}$}&\tabincell{l}{0.9534}&Logistic&\tabincell{l}{$\mu=2.3326\ast 10^{7}$,\\$\sigma=65133.9  $}&\tabincell{l}{0.9987}\\
\hline

Mean  inclination(deg)&Stable&\tabincell{l}{$\alpha = 0.835022 $,\\$\beta =-0.329864$,\\$c=0.242389$,\\$\mu = 165.084$}&\tabincell{l}{0.9076}&$t$ location-scale&\tabincell{l}{$\mu=165.117$,\\$\sigma=0.17015  $,\\$\nu=0.875108$}&\tabincell{l}{0.6662}\\
\hline

Mean  eccentricity&Stable&\tabincell{l}{$\alpha = 1.27463 $,\\$\beta =0.000476987$,\\$c=0.0119451$,\\$\mu = 0.256685 $}&\tabincell{l}{0.9996}&Birnbaum-Saunders&\tabincell{l}{$\beta=0.254254$,\\$\gamma=0.0330888$}&\tabincell{l}{0.6244}\\
 \toprule
\end{tabular}}
\end{table}
\hspace{10mm}\\
\begin{table}[]
 \renewcommand{\arraystretch}{1}
\newcommand{\tabincell}[2]{\begin{tabular}{@{}#1@{}}#2\end{tabular}}
\centering
\caption{\label{opt5}Distributions of current and previous orbital characteristics in the Pasiphae group}
\footnotesize
\rm
\scalebox{0.8}{
\begin{tabular}{@{}*{7}{l}}
 \toprule
&&\textbf{Current distribution}&&&\textbf{Previous distribution}\\
 \toprule
\textbf{Characterastic}&\textbf{Distribution}&\textbf{Parameters}&\textbf{$p$-value}&\textbf{Distribution}&\textbf{Parameters}&\textbf{$p$-value}\\
 \toprule

Semi-major  axis(km)&Extreme Value&\tabincell{l}{$\mu= 2.38737\ast 10^{7}$,\\$\sigma =493901$}&\tabincell{l}{0.9956}&$t$ location-scale&\tabincell{l}{$\mu=3.39242 \ast 10^{7}$,\\$\sigma=273627  $,\\$\nu=0.730397$}&\tabincell{l}{0.3730}\\
\hline

Mean  inclination(deg)&Normal&\tabincell{l}{$\mu =151.213  $,\\$\sigma =4.33208 $}&\tabincell{l}{0.9989}&Logistic&\tabincell{l}{$\mu=151.357$,\\$\sigma=65133.9  $}&\tabincell{l}{0.8987}\\
\hline

Mean  eccentricity&Birnbaum-Saunders&\tabincell{l}{$\beta =0.33437 $,\\$\gamma  =0.24734 $}&\tabincell{l}{0.7280}&Logistic&\tabincell{l}{$\mu=0.295251 $,\\$\sigma=0.0599373   $}&\tabincell{l}{0.9550}\\
 \toprule
\end{tabular}}
\end{table}
In the Pasiphae group, the distributions of orbital features are different from those of the other two groups. As shown in Table 6, the best-fitting distribution of the semi-major axis is an Extreme Value distribution with a $p$-value is 0.9956, which is much larger than the value found in the literature \cite{Gao}. The cause of this phenomenon may be the change in the classification of moons, and more distributions have been tested in this paper than in previous papers.

To more intuitively observe the difference between the current distribution and the previous distribution, the observed cumulative distribution function (CDF) and best-fitting CDF were plotted. Based on the previous data and the new data, the orbital properties of the moons can be compared more specifically and conveniently, and then we obtain the following CDFs (Figures 1 and 2).
\begin{figure}[htbp]
  \centering
  \subfigure[]{\label{fig:subfig:a}
    \includegraphics[width=13cm]{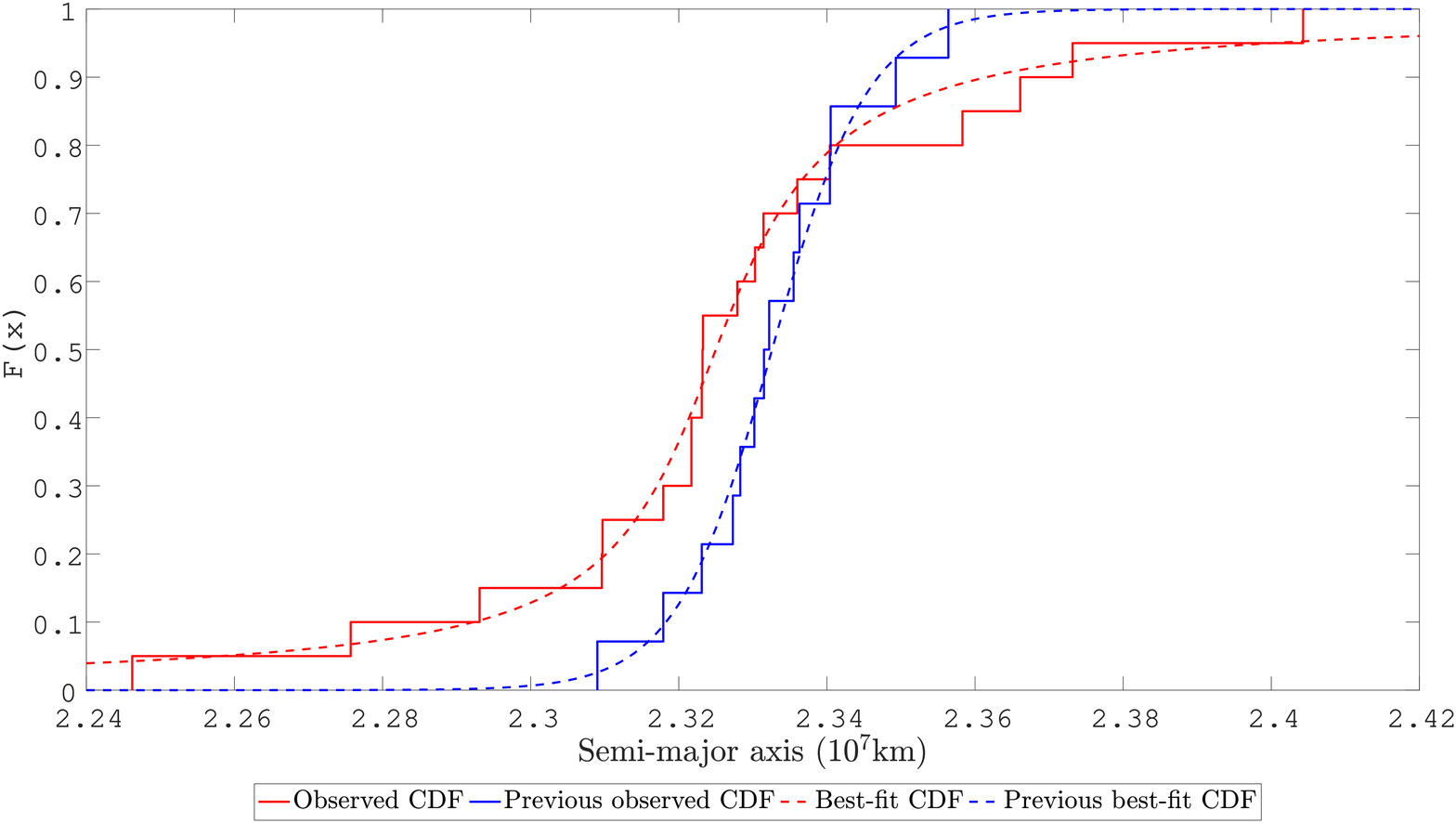}}
  \hspace{1in}
  \subfigure[]{\label{fig:subfig:b}
    \includegraphics[width=13cm]{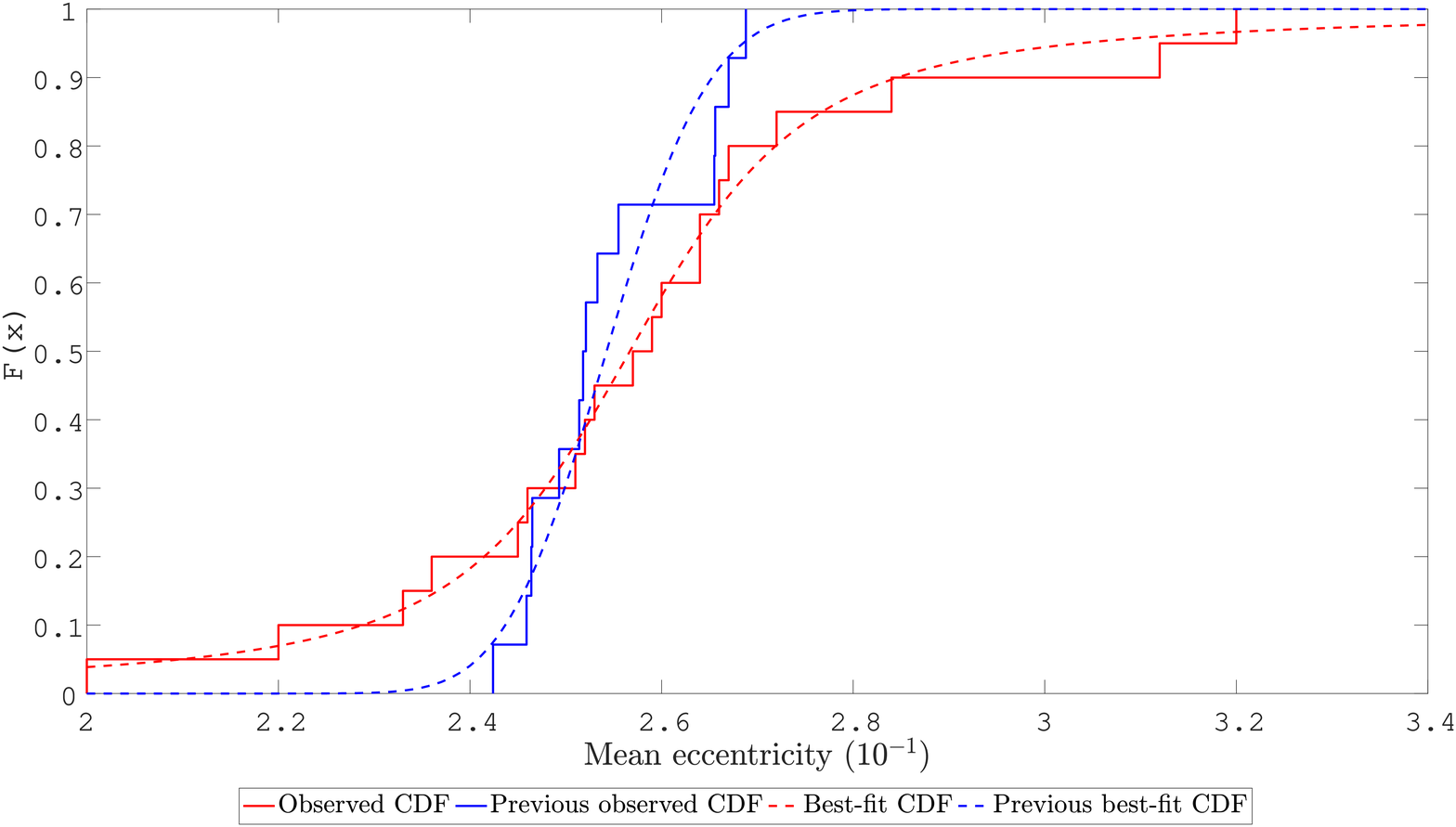}}
     \hspace{1in}
  \subfigure[]{\label{fig:subfig:c}
    \includegraphics[width=13cm]{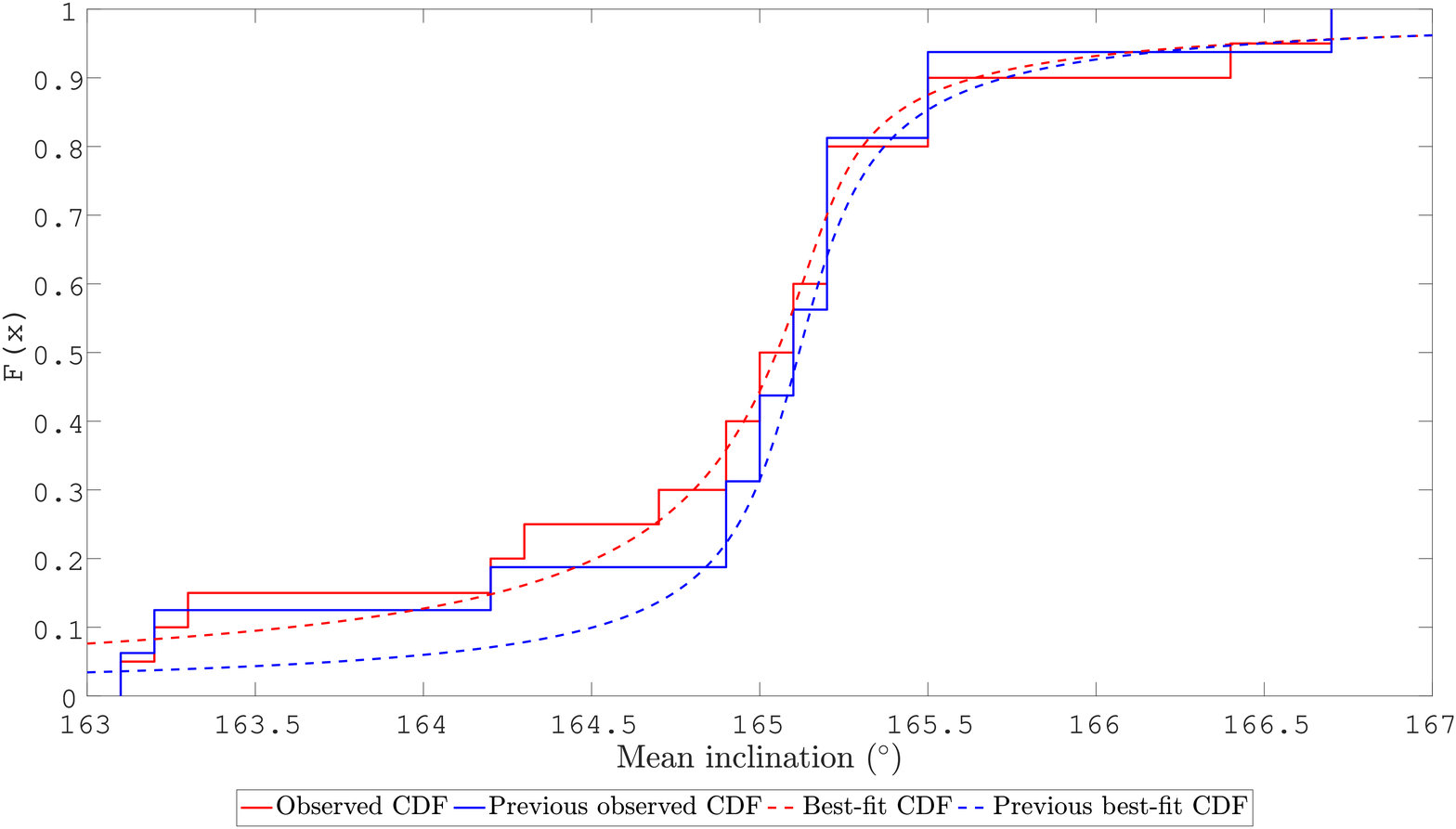}}
  \caption{(a), (b) and (c) are the best-fitting CDF, the observed CDF of the current distributions and the observed CDF of the previous distributions, respectively, of the orbital characteristics of the moons in the Carme group}
  \label{fig:subfig}
\end{figure}
\begin{figure}[htbp]
  \centering
  \subfigure[]{\label{fig:subfig:a1}
    \includegraphics[width=13cm]{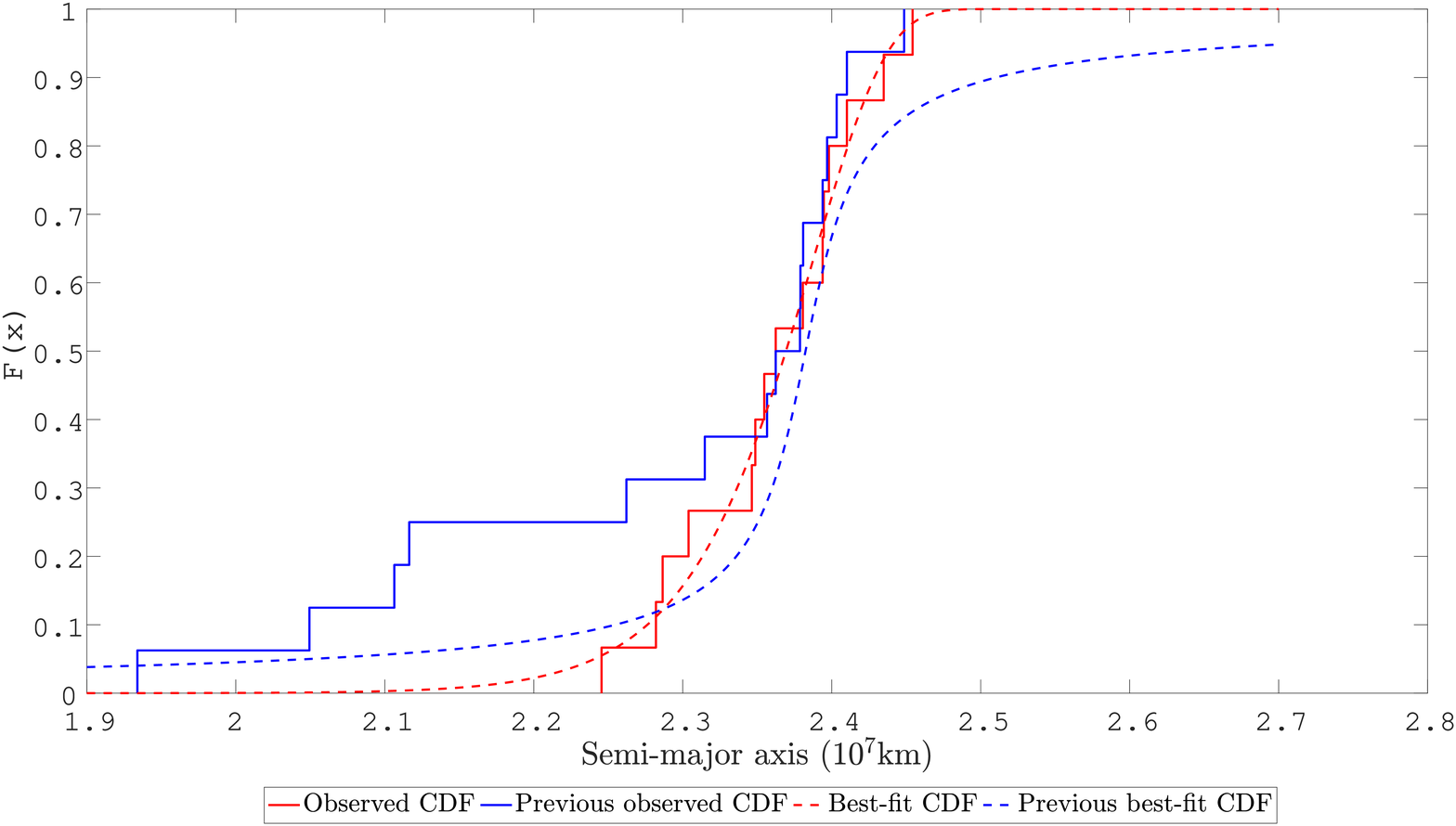}}
  \hspace{1in}
  \subfigure[]{\label{fig:subfig:b2}
    \includegraphics[width=13cm]{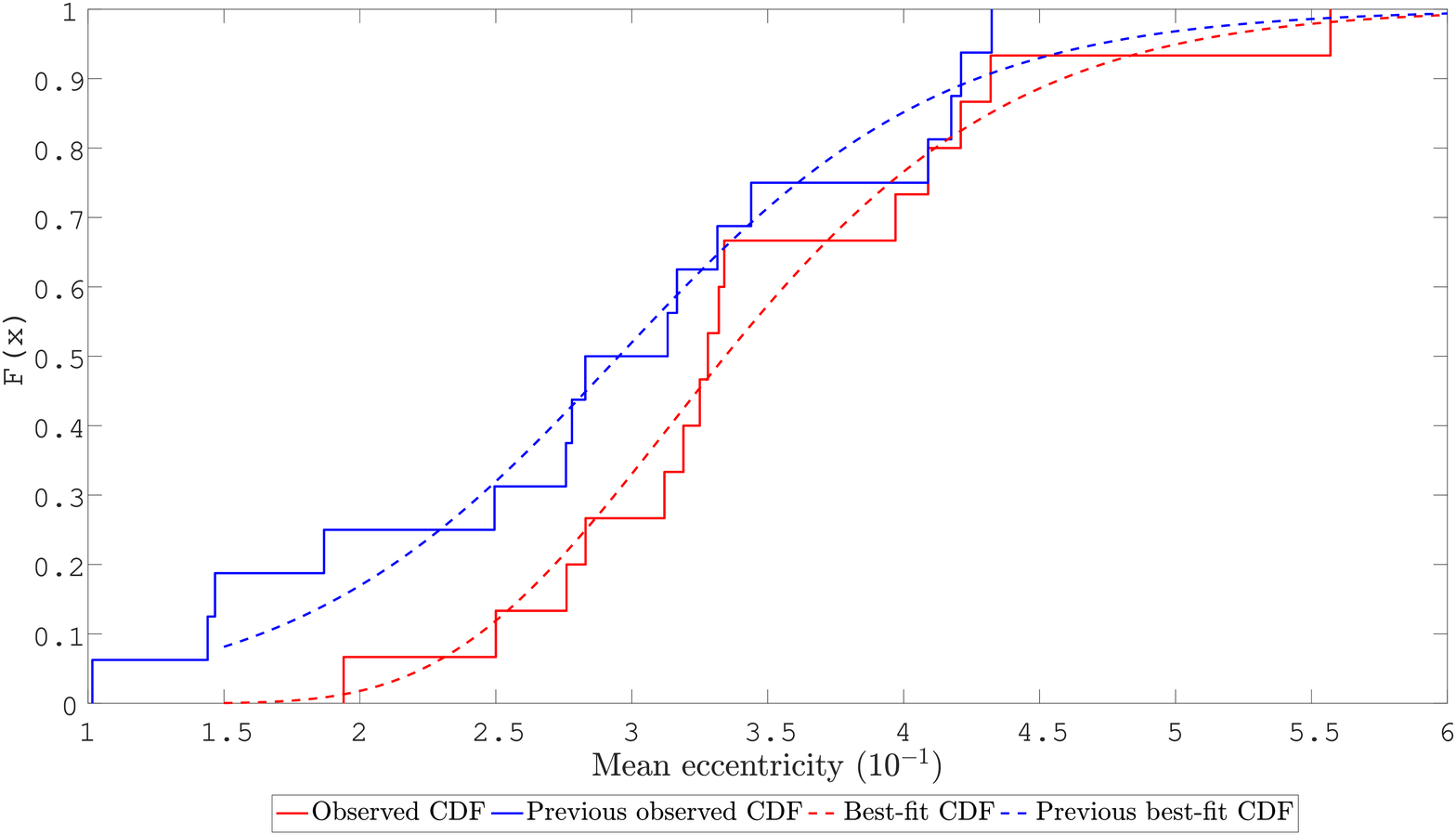}}
     \hspace{1in}
  \subfigure[]{\label{fig:subfig:c3}
    \includegraphics[width=13cm]{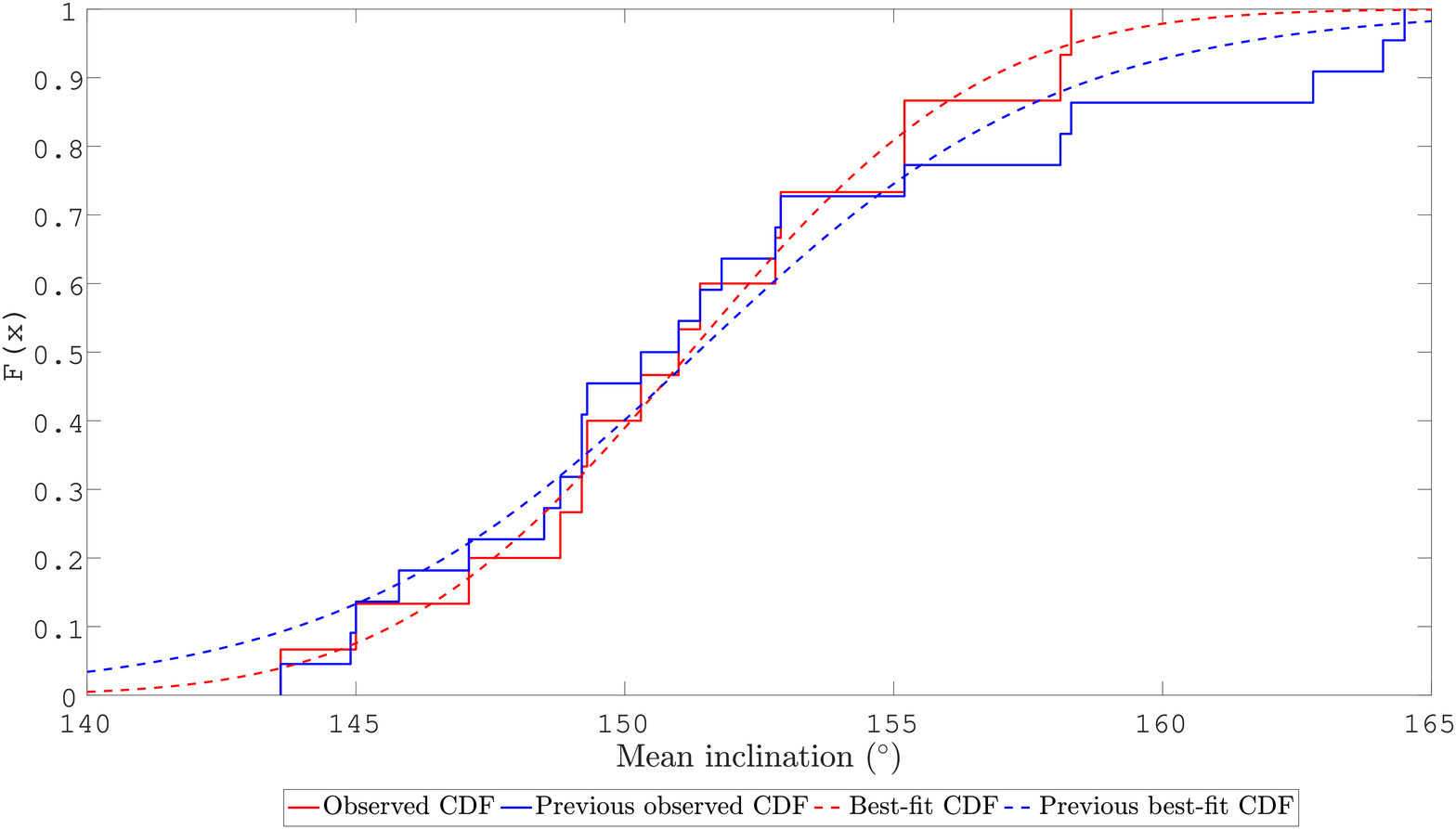}}
  \caption{(a), (b) and (c) are the best-fitting CDF, the observed CDF of the current distributions and the observed CDF of the previous distributions, respectively, of the orbital characteristics of the moons in the Pasiphae group}
  \label{fig:subfig}
\end{figure}

\section{Verification of the rationality of the theoretical results}

In this section, the reasonability of the best-fitting distribution of the semi-major axis and the orbital period is demonstrated analytically based on Kepler's third law of planetary motion. 
\subsection{Carme group}
As seen from Table 2, for either the semi-major axis or the  period, the distribution with the largest $p$-value,  is a Stable distribution. However, as mentioned in Section 2, the probability density function (PDF) of the Stable distribution is given by the characteristic function. Three simple cases, namely, a Gaussian distribution, Cauchy distribution and Levy distribution, are not occurred, and it is difficult to calculate and draw the PDF. Therefore, we discuss the second largest $p$-value $t$ location-scale distribution. When the semi-major axis obeys the $t$ location-scale distribution, the $p$-value is 0.9000, and the parameters are 2.32508, 0.01141172, and 1.39731. When the period obeys a $t$ location-scale distribution, the $p$-value is 0.9792, and the parameters are 724.436, 6.75988, and 1.01233. The pdf of the $t$ location-scale distribution is given.

It can be seen from Table 2 that both the semi-major axis and the period obey a Stable distribution, and according to Kepler's third law, there is a relationship $T=\sqrt{4\pi ^{2}a^{3}/GM}$  ($G$ is the universal gravitational constant, and $M$ is the mass of Jupiter) between them. In theory, if the distribution type of the semi-major axis or the period is known, then the other one can be derived analytically. If the analysis results are consistent with the statistical inference results, or are at least very close, then the statistical inference results are valid. However, here, there is a problem, as described in Section 2, and the Stable distribution of the PDF can be given by the CF. The three special cases, namely, the Gaussian distribution, Cauchy distribution and Levy distribution are currently well studied, but in other Stable distributions, the PDFs are still poorly studied. Therefore, we have to discuss the distribution that is very close to its $p$-value, i.e. the $t$ location-scale distribution. When the semi-major axis obeys a $t$ location-scale distribution with a $p$-value of 0.9000, the corresponding parameters are 2.32508, 0.01141172, and 1.397731. When the period obeys a $t$ location-scale distribution of the $p$-value of 0.9792, the corresponding parameters are 724.436, 6.75988, and 1.01233.

Note that the PDF of the $t$ location-scale distribution can be defined as
\begin{equation}
\begin{aligned}
f= &\frac{\Gamma (\frac{\nu +1}{2}))}{\sigma \sqrt{\nu \pi }\Gamma (\frac{\nu }{2}))}\left [\frac{\nu +(\frac{x-\mu }{\sigma })^{2}}{\nu }  \right ]^{-\frac{\nu +1}{2}} ,
\end{aligned}
\end{equation}
where $\Gamma (\cdot )$ is the gamma function.

Therefore, the corresponding predicted PDFs can be written as
\begin{equation}
\begin{aligned}
f_{pre,\,a}(a;\mu ,\sigma ,\nu )= \frac{\Gamma (\frac{\nu +1}{2}))}{\sigma \sqrt{\nu \pi }\Gamma (\frac{\nu }{2}))}\left [\frac{\nu +(\frac{a-\mu }{\sigma })^{2}}{\nu }  \right ]^{-\frac{\nu +1}{2}}
=\frac{23.884}{[1+0.716(70.836a-164.698)^{2}]^{1.199}},
\end{aligned}
\end{equation}
and
\begin{equation}
\begin{aligned}
f_{pre,\,T}(T;\mu ,\sigma ,\nu )= \frac{\Gamma (\frac{\nu +1}{2}))}{\sigma \sqrt{\nu \pi }\Gamma (\frac{\nu }{2}))}\left [\frac{\nu +(\frac{T-\mu }{\sigma })^{2}}{\nu }  \right ]^{-\frac{\nu +1}{2}}
=\frac{0.047}{[1+0.988(0.148a-107.181)^{2}]^{1.006}}.
\end{aligned}
\end{equation}
Then, based on Kepler's third law, the PDF of the semi-major axis can be derived as follows:
\begin{equation}
\begin{aligned}
f_{ana,a}(T;\mu ,\sigma ,\nu )&=3\pi \sqrt{\frac{a}{GM}}f_{pre,T}(\sqrt{\frac{4\pi ^{2}a^3}{GM}};724.436,6.75898,1.01233)\\
&=\frac{14.475\sqrt{a}}{[1+0.988(30.245\sqrt{a^{3}}-107.181)^{2}]^{1.006}}.
\end{aligned}
\end{equation}

From Figure 3 (a), we can find that $f_{ana,\,a}(T;\mu ,\sigma ,\nu )$ (the PDF is represented by the red curve) obtained by the analytical method is very similar to $f_{pre,\,a}(a;\mu ,\sigma ,\nu )$(the PDF is represented by the blue curve) obtained by statistical inference.
\begin{figure}[]
  \centering
  \begin{minipage}{180mm}
  \subfigure[]{\label{fig:subfig:a}
    \includegraphics[width=9cm]{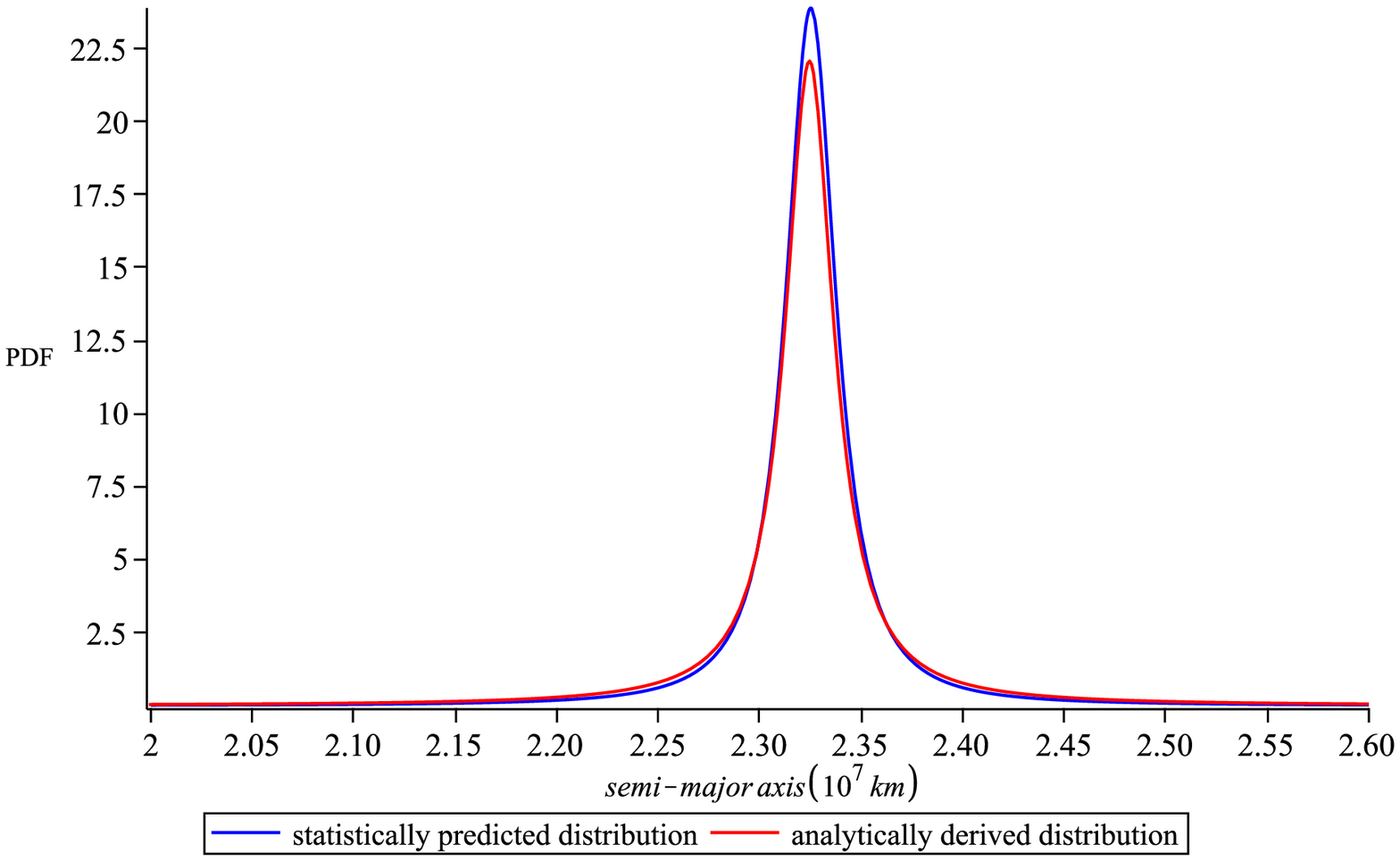}}
  \subfigure[]{\label{fig:subfig:b}
    \includegraphics[width=9cm]{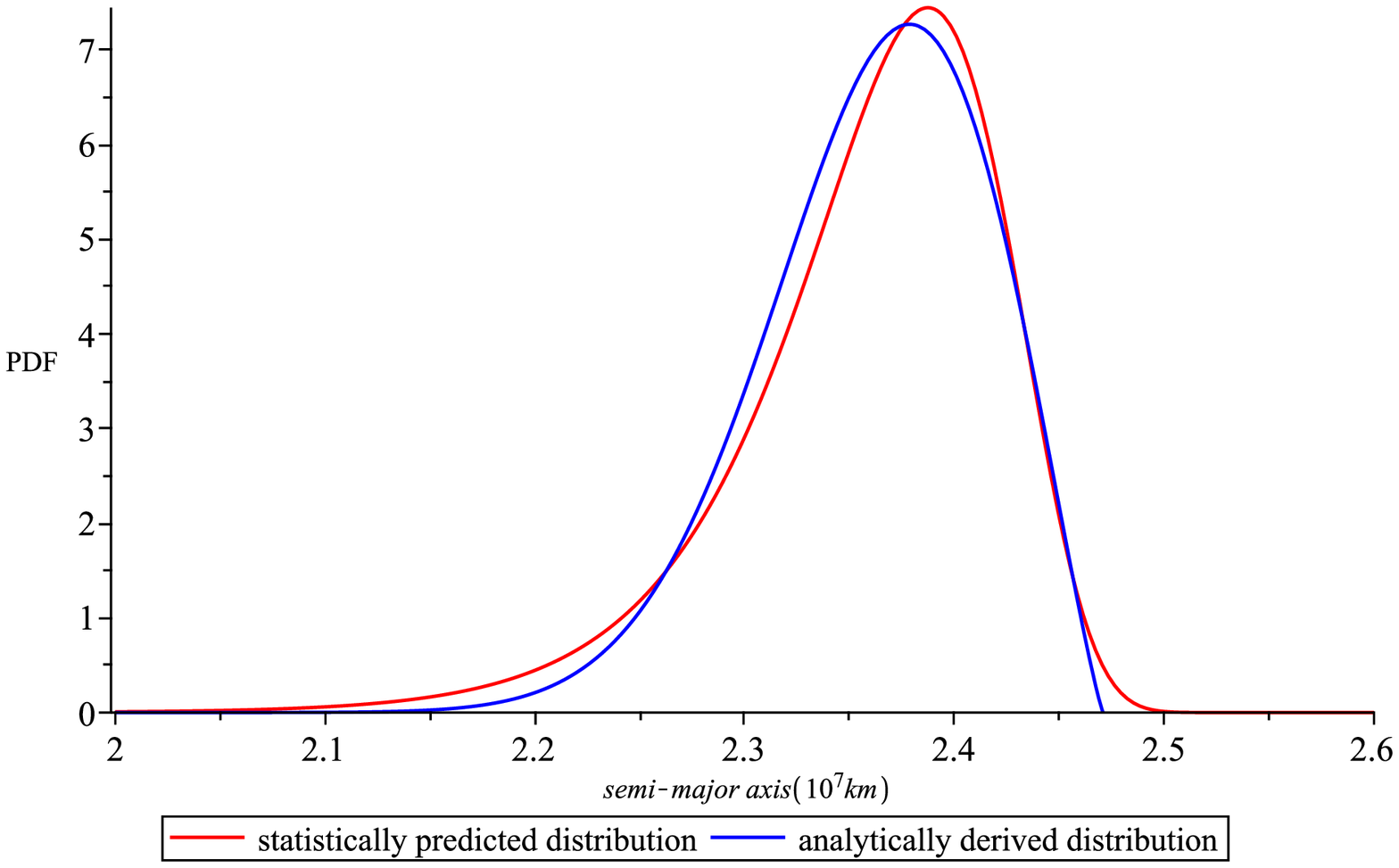}}
  \caption{(a) and (b) are the PDFs of semi-major axis in the Carme group and the Pasiphae group, respectively}
  \label{fig:subfig}
  \end{minipage}
 \end{figure}
\subsection{Pasiphae group}
The corresponding predicted PDFs can be denoted by
\begin{equation}
\begin{aligned}
f_{pre,\,a}(a;\mu ,\sigma )=\frac{e^{-\frac{a-\mu }{\sigma }}e^{-e^{\frac{a-\mu }{\sigma }}}}{\sigma }
=20.247e^{20.247a-48.337}e^{-e^{20.247a-48.337}},
\end{aligned}
\end{equation}
and
\begin{equation}
\begin{aligned}
f_{pre,\,T}(T;k,\mu ,\sigma )
&=\frac{e^{-(1+((T-\mu )/\sigma) ^{-1/k})}(1+k((T-\mu )/\sigma )^{-1-1/k})}{\sigma }\\
&=0.371e^{-(13.458-0.017T)^{2.171}}(13.458-0.017T)^{1.171}.
\end{aligned}
\end{equation}
Similarly, by using Kepler's third law, the PDF of the semi-major axis can also be derived analytically as follows
\begin{equation}
\begin{aligned}
f_{ana,\,a}(T;\mu ,\sigma )
&=3\pi\sqrt{\frac{a}{GM}}f_{pre,T}(\sqrt{\frac{4\pi ^{2}a^{3}}{GM}};-0.460585,26.9631,729.292)\\
&=11.284\sqrt{a}e^{-(13.458-3.465\sqrt{a^{3}})^{2.171}}(13.458-1.103\pi\sqrt{a^{3}})^{1.171}\\
\end{aligned}
\end{equation}

From Figure 3 (b), we can also find that $f_{ana,\,a}(T;\mu ,\sigma )$ (the PDF is represented by the blue curve) obtained by the analytical method is in good agreement with $f_{pre,\,T}(T;k,\mu ,\sigma )$ (the PDF is represented by the red curve) obtained by statistical inference.

\section{Conclusions}
Based on reference \cite{Gao}, we use the K-S test to study the distributions of the orbital elements and orbital periods of the most recently discovered Jovial irregular moons in this paper. These orbital features mainly obey Stable, Extreme Value, Loglogistic, Generalized Extreme Value, Normal and Birnbaum-Saunders distributions.

Moreover, we also made some comparisons on the distributions of the semi-major axis, the mean eccentricity and the mean inclination. From the comparison results, the best-fitting distribution of the three features in this paper has the larger $p$-value. From the figures of best-fitting CDF and the CDF based on the observational data, the current best-fitting distribution and the previous one are well matched. There are two possible reasons for this result. First, the number of tested distribution functions is greater than that in \cite{Gao}. Second, the classification of Jupiter's moons has changed, and 12 newly discovered moons have been added. Furthermore, based on Kepler's third law, the PDF obtained by the analytical method is very close to the PDF obtained by statistical inference, so it is reasonable to say that the best-fitting distribution of these orbital features is reasonable.

In addition, Table 5 shows that the orbital elements of some moons have the same Stable distribution. This interesting result may indicate that they have the same origin; they may have originated from the same parent asteroid. We will continue to study whether they are truly `siblings'.

\section*{Appendix}

\textbf{A. Classification of Irregular Moons}

see Table 7

\textbf{B. Orbital characteristics of Irregular Moons}

see Table 8

\textbf{C. Distribution Inference Results}

See Tables 9-17

\clearpage
\begin{table}[]
\footnotesize
\centering
\caption{Classification of irregular moons}
\begin{tabular}{llllll}
\midrule
\midrule
\textbf{Number} 
& \textbf{Name}  
&\textbf{Designation}
& \textbf{Number}
& \textbf{Name}  
& \textbf{Designation} \\
 \toprule
\multicolumn{3}{c}{\textbf{Ananke group (Current)}} & \multicolumn{3}{c}{\textbf{Ananke group (Previous)}} \\
 \toprule
XXXIV & Euporie & \multicolumn{1}{c}{S/2001 J10} &&&  \\
LV&& \multicolumn{1}{c}{S/2003 J18} &&&  \\
LII&& \multicolumn{1}{c}{S/2010 J2} & LII&& S/2010 J2 \\
&& \multicolumn{1}{c}{S/2017 J7} &&&  \\
LIV&& \multicolumn{1}{c}{S/2016 J1} &&&  \\
&& \multicolumn{1}{c}{S/2017 J3} &&&  \\
\end{tabular}
\end{table}

\clearpage
\begin{table}[htbp]
  \centering
  \caption{Orbital characteristics of irregular moons in Ananke, Carme and Pasiphae groups$^1$}
\scalebox{0.6}{    
\begin{tabular}{ccccccccc}
    \toprule
    \toprule
    \multicolumn{9}{c}{\textbf{Ananke group}} \\
    \midrule
    \midrule
    \textbf{Number} & \textbf{Name } & \textbf{Designation} & \textbf{Mean semi-major axis (km)} & \textbf{Mean inclination (deg)} & \textbf{Mean eccentricity} & \textbf{Peri (deg)} & \textbf{Node (deg)} & \textbf{Period (days)} \\
    \midrule
    XXXIV & Euporie & S/2001 J10 & 19302000 & 145.8 & 0.144 & 74.6  & 64.9  & 550.7 \\
    LV    &       & S/2003 J18 & 20274000 & 146.4 & 0.105 & 98.15 & 215.5 & 588 \\
    LII   &       & S/2010 J2 & 20307150 & 150.4 & 0.307 & 0     & 0     & 588.1 \\
          &       & S/2017 J7 & 20627000 & 143.4 & 0.215 & 323.5 & 321.5 & 602.6 \\
    LIV   &       & S/2016 J1 & 20650845 & 139.8 & 0.141 & 328.2 & 293.8 & 602.7 \\
          &       & S/2017 J3 & 20694000 & 147.9 & 0.148 & 171.6 & 82.5  & 606.3 \\
\bottomrule
    \bottomrule
    \end{tabular}%
}
  \label{tab:addlabel}%
\end{table}%
\footnotetext[1]{see https://sites.google.com/carnegiescience.edu/sheppard/moons/jupitermoons for more details.}

\begin{landscape}
\begin{table}[]
 \renewcommand{\arraystretch}{1.2}
\tiny
\centering
\caption{Ananke group}
\scalebox{0.88}{
%
}
\end{table}
\end{landscape}
\begin{landscape}
\begin{table}[]
 \renewcommand{\arraystretch}{1.2}
\tiny
\centering
\caption{Ananke group (Cont'd)}
\scalebox{0.88}{
%
}
\end{table}
\end{landscape}

\begin{landscape}
\begin{table}[]
 \renewcommand{\arraystretch}{1.2}
\tiny
\centering
\caption{Ananke group (Cont'd)}
\scalebox{0.88}{
%
}
\end{table}
\end{landscape}

\begin{landscape}
\begin{table}[]
 \renewcommand{\arraystretch}{1.2}
\tiny
\centering
\caption{Carme group}
\scalebox{0.88}{
%
}
\end{table}
\end{landscape}

\begin{landscape}
\begin{table}[]
 \renewcommand{\arraystretch}{1.2}\tiny
\centering
\caption{Carme group (Cont'd)}
\scalebox{0.88}{
%
}
\end{table}
\end{landscape}

\begin{landscape}
\begin{table}[]
 \renewcommand{\arraystretch}{1.2}\tiny
\centering
\caption{Carme group (Cont'd)}
\scalebox{0.88}{
%
}
\end{table}
\end{landscape}

 \begin{landscape}
\begin{table}[]
 \renewcommand{\arraystretch}{1.2}\tiny
\centering
\caption{Pasiphae group}
\scalebox{0.88}{
%
}
\end{table}
\end{landscape}

\begin{landscape}
\begin{table}[]
 \renewcommand{\arraystretch}{1.2}\tiny
\centering
\caption{Pasiphae group (Cont'd)}
\scalebox{0.88}{
%
}
\end{table}
\end{landscape}

\begin{landscape}
\begin{table}[]
 \renewcommand{\arraystretch}{1.2}
\tiny
\centering
\caption{Pasiphae group (Cont'd)}
\scalebox{0.88}{
%
  \\
\bottomrule
\end{tabular}}
\end{table}
\end{landscape}

\clearpage

 \noindent \textbf{Author Contributions}\\
Formal analysis, Fabao Gao; Software, Xia Liu; Writing---original draft, Xia Liu; Writing---review $\&$ editing, Fabao Gao.\\
 
\noindent  \textbf{Funding}\\
This research was funded by the National Natural Science Foundation of China (NSFC) though grant No.11672259 and the China Scholarship Council through grant No.201908320086.\\

\noindent  \textbf{Conflicts of Interest}\\
The authors declare that there is no competing interests.

\section*{Acknowledgments}
We are very grateful to Dr. Abedin Y. Abedin for his comments and suggestions which help us to improve the presentation of this paper greatly.

\end{document}